\documentclass[a4paper]{article}
\usepackage{RR}
\usepackage{hyperref}

\usepackage{amsmath}
\usepackage{graphicx}
\usepackage{color}
\usepackage{float}
\usepackage{rotating}
\usepackage{cite}
\usepackage{listings}

\lstdefinestyle{custom_code}
	{numbers=left,columns=fullflexible, commentstyle=\textit}
\lstdefinestyle{example_code}
		{numbers=left,columns=fullflexible, commentstyle=\textit}
\lstdefinestyle{classification_code}
		{numbers=none,columns=fullflexible, commentstyle=\textit}
\lstdefinestyle{numerable_code}
		{numbers=left,columns=fullflexible, commentstyle=\textit}
		
\RRNo{6501}		
\RRdate{March 2008}
\RRauthor{
Freddy Munoz%
  \and
Benoit Baudry%
  \and
Olivier Barais%
}

\authorhead{Munoz \& Baudry \& Barais}

\RRtitle{}

\RRetitle{A classification of invasive patterns in AOP}

\titlehead{A classification of invasive patterns in AOP}

\RRabstract{
Aspect-Oriented Programming (AOP) improves modularity by encapsulating crosscutting concerns into aspects. Some mechanisms to compose aspects allow invasiveness as a mean to integrate concerns. Invasiveness means that AOP languages have unrestricted access to program properties. Such kind of languages are interesting because they allow performing complex operations and better introduce functionalities. In this report we present a classification of invasive patterns in AOP. This classification characterizes the aspects invasive behavior and allows developers to abstract about the aspect incidence over the program they crosscut.
}

\RRkeyword{Aspect-oriented programming, Classification system, Invasive AOP.}

\RRprojets{Triskell}

\RRtheme{\THCom}

\URRennes

\begin{document}
	
	\makeRR

	\section{Introduction}

Aspect-oriented Programming (AOP) is a paradigm that enhances current approaches to modularizing software. AOP enables separation of concerns that crosscut the implementation of a system. This is done by encapsulating crosscutting concerns into single units called \emph{aspects}. An aspect itself is composed of several units that realize the crosscutting behavior. Aspects also provide pointing elements that designate well defined points in the program execution or structure. These are the points where the program executes the crosscutting behavior. Different approaches to AOP have been proposed~\cite{Kiczales:1997:AOP,643613,643606,conf/cbse/SuveeFV06,1101566,Bockisch06}. Each of these approaches provide a different mechanism to compose aspects with the base program. These composition mechanisms range from simple program augmentation to more complex operations such as behavior replacement. 
		
\emph{Invasive} composition mechanism allow developers to manipulate almost any structure and behavior of the base program. \emph{Invasive AOP} approaches realize invasive composition mechanism. \emph{Invasiveness} is the ability of invasive AOP to manipulate the program structures and behavior. Invasive AOP provides several strategies to manipulate the base program. These strategies range from less invasive such as the augmentation of a procedure execution to more invasive ones such as the replacement of a procedure execution.
	
This report presents a classification system for invasive AOP. This classification system focuses on the invasive behavior of aspects over the base program.  We construct this classification system by studying the invasive structures of AspectJ~\cite{aspectj/lr}, however, we think that it can be adapted to other invasive languages.
	
	The reminder of this report is organized as follows. In section 2 we study the AspectJ structures in order to derive our classification of invasive AOP. In section 3 we introduce our classification of invasive aspects. In section 4 we discuss related work. Finally section 5 concludes.

	%
	%
	%
	%
	%
	%
	%
	%

\section{Classifying Invasive AOP}

	Each programming paradigm promotes its own way to encapsulate concerns. For example, the object-oriented paradigm encapsulates concerns in classes, fields and methods. It also defines a protection level for each one of them. This ensures that properties will be accessed according to their permission.

	Invasiveness breaks the encapsulation that each paradigm promotes and provides a mean to access protected properties. For example, aspects can access classes, methods and fields whatever their permission. Developers can fruitfully use invasiveness. For example, they can use it to check the preservation of system properties~\cite{Wampler:aosd-acp4is06}. However, developers can use invasiveness to invalidate some desirable system properties, intentionally or unintentionally.
	
	The duality in the usage of invasiveness motivates us to propose a classification system for invasive AOP. Our classification system provides the following benefits.

	\begin{itemize}
		\item{It enables the characterization of aspects with precise invasive behavior (invasive pattern).}
		\item{It enables developers to reason about aspects and discriminate when they could represent a risk to the base program.}
	\end{itemize}
 	
	We build the classification system by studying the invasive structures available in AspectJ. Once identified the different possibilities it provides, we proceed to derive our classification system.

\subsection{Running example}

	To illustrate our characterization approach, in listing~\ref{code:example} we present the skeleton implementation of an ordered list class (\texttt{MyArrayList}). All the attributes in this list are private to the class. Elements added using the \texttt{add} method (line 8) are stores in the \texttt{elements} array (line 2). Elements are retrieved by using the \texttt{get} method (line 11). The specific position of an element in the \texttt{elements} array is obtained by using the \texttt{indexOf} method (line 12). An element is removed from the list by using the \texttt{remove} method (lines 13-14), this method removes an element by using its position or the original element. The first and last element in the list are retrieved by using the methods \texttt{first} and \texttt{last} (lines 15,16). All the elements in the list can be removed by using the \texttt{clear} method (line 9).
	
	\lstset{language=[AspectJ]Java}
	\lstset{style=classification_code}
\begin{lstlisting}[numbers=left,caption=Skeleton implementation of an ordered list example,captionpos=b,label=code:example]
public class MyArrayList {
	private Object[] elements;
	private int initialSize=0;
	private int growFactor=5;
	private int size=0;
	public MyArrayList(int grow,int initialSizei){...}
	public MyArrayList(){..}
	public boolean add(Object o) {...}
	public void clear() {...}
	public boolean contains(Object o) {...}
	public Object get(int index) {...}
	public int indexOf(Object o) {...}
	public boolean remove(Object o) {...}
	public Object remove(int index) {...}
	public Object first() {...}
	public Object last() {...}
	public int size() {...}
}
\end{lstlisting}
		
	Listing~\ref{code:aspect} presents the code of our example aspect. This aspect (\texttt{MyAspect}) realizes a set of advices an Inter-type declarations that affect in different ways the \texttt{MyArrayList} class. The declaration in line 3 introduces the parent interface \texttt{Comparable} into the \texttt{MyArrayList} class hierarchy. The inter-type in line 5 adds the method \texttt{compareTo} to the  \texttt{MyArrayList} class. Analogously, the inter-type of line 7 adds the field \texttt{maxSize}. The advice in lines 9-12 execute a logging facility just before the execution of the \texttt{add} method. The advice in lines 14-19 print a message into the standard output before and after the execution of the \texttt{add} method. The advice in lines 21-24 replaces the execution of the \texttt{first} method and executes the \texttt{last} method instead. The advice in lines 26-32 executes the \texttt{add} method only when non null values are passed as parameters. The advice in lines 34-42 adds each element of a collection into the list. The advice in lines 44-48 replaces the parameter object (object to be inserted) by an string, then the string is inserted instead of the object. The advice in lines 48-52 reads the field current size and print it to the standard output. Finally, the advice in lines 54-57 decreases the value of \texttt{currentSize} by 1.

	\lstset{style=classification_code}
\begin{lstlisting}[numbers=left,caption=MyAspect example,captionpos=b,label=code:aspect]
public privileged aspect MyAspect{
	
	declare parents: MyArrayList implements Comparable;
	
	public boolean MyArrayList.compareTo(Object o){...}
	
	public int MyArrayList.maxSize=100;
	
	before(MyArrayList list,Object o):
		execution(* MyArrayList+.add(Object+)) && args(o) && this(list){
		logger.debug("Adding object "+o.toString());
	}
	
	void around((MyArrayList list,Object o):execution(* MyArrayList+.add(Object)) &&
		args(obj) && target(list){
		System.out.println("adding object:"+obj);
		proceed(obj,list);
		System.out.println("object added ");
	}
	
	boolean around(MyArrayList list):
		execution(* MyArrayList+.first()) && this(list){
		return list.last();
	}
	
	boolean around(MyArrayList list,Object o):
		execution(* MyArrayList+.add(Object+)) && args(o) && this(list){
		if(o!=null){ 
		  return proceed(list,o); 
		}
		return false;
	}
	
	boolean around(MyArrayList list,Collection c):
		call(* MyArrayList+.add(Collection+)) && args(c) && this(list){
		for(Object o:c){ 
		   proceed(list,o); 
		}
		return true;
	}
	
	boolean around(MyList list,Object o):
		call(* MyArrayList+.add(Object+)) && args(o) && this(list){
		o = new String("object"+ o.toString());
		return proceed(list,o);
	}

	after(MyArrayList list):
		execution(* MyArrayList+.add(..)) && this(list){
		int size=list.currentSize;
		logger.debug("current size: "+size);
	}
	
	before(MyArrayList list):
		execution(* MyArrayList+.add(..)) && this(list){
		list.currentSize=list.currentSize-1;
	}
	
	boolean around(MyArrayList list,int i):
		call(* MyArrayList+.get(int)) && args(i) && this(list){
		Object o=proceed(list,i);
		return o.toString();
	}
}
\end{lstlisting}

\subsection{Invasive AOP: The case of AspectJ}

	AspectJ~\cite{aspectj/lr} is the most prominent realization of invasive aspects. It realizes the crosscutting behavior on \emph{advices} and designating the places where this behavior must be woven with pointcut expressions. Pointcuts are regular expressions that match well defined point in the structure (static) and execution (dynamic) of the program.
	
	In AspectJ, aspects are composed of advices, pointcuts, inter-type declarations and inter-type parent declarations. Advices are used to modify the program flow and write or read fields. An advice can be declared inside a privileged aspect, which means that it is enabled to ignore the object-oriented access policy. Inter-type declarations are used to introduce methods and fields into a target class. Inter-type parent declarations are declarations of inheritance that modify the class hierarchy.

	AspectJ allows developers to implement various behaviors on advices. We summarize them by using the actions advices may perform in the following list.
	
	\begin{itemize}
		\item{The advice does not modify the behavior of the methods it crosscuts. (Listing~\ref{code:aspect}, lines 9-12, 14-19)}
		\item{The advice replaces the behavior of the methods it crosscuts, the original methods are never invoked. (Listing~\ref{code:aspect}, lines 21-24)}
		\item{The advice conditionally replaces the behavior of the methods it crosscuts. (Listing~\ref{code:aspect}, lines 26-32)}
		\item{The advice replaces the behavior of the methods it crosscuts. It invokes two or more time the original method. (Listing~\ref{code:aspect}, lines 34-40)}
		\item{The advice invokes a method declared outside its declaring aspect. (Listing~\ref{code:aspect}, line 11)}
		\item{The advice changes the argument values of the methods it crosscuts. (Listing~\ref{code:aspect}, lines 42-46)}
		\item{The advice replaces the result of the method execution. (Listing~\ref{code:aspect}, lines 59-63}
		\item{The advice reads or writes object fields values. (Listing~\ref{code:aspect}, lines 50, 56)}
	\end{itemize}

	AspectJ allows developers to modify the program structure. We summarize the modifications developers can implement on aspects in the following list. It is worth to remark that these modifications are performed at the aspect level.
	
	\begin{itemize}
		\item{The aspect inserts new fields into a target class by using an inter-type declaration. (Listing~\ref{code:aspect}, line 7)}
		\item{The aspect inserts new methods into a target class by using an inter-type declaration. (Listing~\ref{code:aspect}, line 5)}
		\item{The aspect modifies the class hierarchy by using an inter-type parent declaration. (Listing~\ref{code:aspect}, line 3)}
	\end{itemize}

	These list of interventions range from simple augmentation to more invasive ones like complete replacing the target method or write a protected field. We motivate our classification on these factual intervention we proceed to identify different invasive behavior that we call \emph{invasiveness patterns}.

	\section{Classification System}
	\label{subsection:aspect_classification}
	
	We can identify three axes for our classification. The first corresponds to how developers can use advices to affect the program control flow. The second corresponds to how developers can use advices to affect the data flow, changing fields, methods parameters and returning values. The third corresponds to how developers can use aspects to modify the program structure.
	
	\subsubsection{Control Flow} Classifies advices according to the behavioral interaction between them and the methods it crosscuts.

	\begin{description}
		\item \emph{Augmentation} : After crosscutting, the body of the method is always executed. The advice augments the behavior of the method it crosscuts with new behavior that does not interfere with the original behavior. The advices in Listing~\ref{code:aspect}, lines 9-12, 14-19 realize the invasiveness pattern \emph{Augmentation}.

		\item \emph{Replacement} : After crosscutting, the body of the method is never executed. The advice completely replaces the behavior of the method it crosscuts with new behavior. This kind of advices eliminate a part of the base code. The advice in Listing~\ref{code:aspect}, lines 21-24 realizes the invasiveness pattern \emph{Replacement}. The advice in lines 59-63 is an special case of \emph{Replacement}, this because it executes the original behavior to later replace the obtained result.

		\item \emph{Conditional replacement} : After crosscutting, the body of the method is not always executed. The advice conditionally invokes the body of the method and potentially replaces its behavior with new behavior. Examples of this kind of advices are advices realizing transaction, access control, etc. Listing~\ref{code:aspect}, lines 26-32 realizes the invasiveness pattern \emph{Conditional replacement}.

		\item \emph{Multiple} : After crosscutting, the body of the method is executed more than once. The advice invokes two or more time the body of the method it crosscuts generating potentially new behavior. The advice in Listing~\ref{code:aspect}, lines 34-40 realizes the invasiveness pattern \emph{Multiple}.

		\item \emph{Crossing} : After crosscutting, the advice invokes the body of a method (or several methods) that it does not crosscut. The advice have a dependency to the class owning the invoked method(s). The advice in Listing~\ref{code:aspect}, lines 9-12 realizes the invasiveness pattern \emph{Crossing}.

	\end{description}

	\subsubsection{Data Access} Classifies advices according to the access they perform to object fields and method arguments.

	\begin{description}	

		\item \emph{Write} : After crosscutting, the advice writes an object field. This access breaks the protection declared for the field and can modify the behavior of the underlying computation. The advice in Listing~\ref{code:aspect}, lines 54-57 realizes the invasiveness pattern \emph{Write}.

		\item \emph{Read} : After crosscutting, the advice reads an object field. This access breaks the protection declared for the field and can potentially expose sensitive data. The advice in Listing~\ref{code:aspect}, lines 48-52 realizes the invasiveness pattern \emph{Read}.

		\item \emph{Argument passing} : After crosscutting, the advice modifies the arguments of the method it crosscuts and then invokes the body of the method. The body of the method always executes at least once. The advice in Listing~\ref{code:aspect}, lines 42-46 realizes the invasiveness pattern \emph{Argument passing}.

	\end{description}

	\subsubsection{Structural}  Classifies aspects according to the modification performed they perform to the existing structure of a class.

	\begin{description}

		\item \emph{Hierarchy} : The aspect modifies the declared class hierarchy. For example, the aspect adds a new parent interface to an existing one. An example of this is the declaration in Listing~\ref{code:aspect}, line 3.

		\item \emph{Field addition} : The aspect adds new fields to an existing class declaration. These fields depending on their protection can be acceded by referencing an object instance of the affected class. An example of this is the declaration in Listing~\ref{code:aspect}, line 7.

		\item \emph{Operation addition} : The aspect adds new methods to an exiting class declaration. These methods depending on their protection can be acceded by referencing an object instance of the affected class. An example of this is the declaration in Listing~\ref{code:aspect}, line 5.

	\end{description}

	The groups we have presented describe the aspect invasive behavior in different dimensions, and most of them are complementary. For example, an advice cannot perform data access without executing and therefore a Write advice can also realize and Augmentation advice. However, some classification categories are incompatible, \emph{Augmentation}, \emph{Replacement} and \emph{Conditional Replacement} cannot co-exist in the same advice. 

	Note that this classification is still incomplete. It does not address cases like the exceptions raised by advices. However, it allows us to characterize the behavioral interaction between advices and methods, major structure modifications and data access.
	
	%
	%
	%
	%
	%
	%
	%
	%
	\section{Related work}

	In~\cite{rinard:fse04} categories of direct and indirect interactions between aspects and methods are identified. Direct interaction is whether an advice interferes with the execution of a method, whereas indirect is whether advices and methods may read/write the same fields. This classification is similar to ours, however, it addresses a different dimension. We identify invasiveness patterns instead of direct/indirect interactions.
 
	%
	Katz~\cite{katz:aosd-foal04} recognizes the fact that aspects can be harmful to the base code and the need of specification on aspect-oriented applications. Our approach agrees with his ideas and likewise we propose a mean to write such specifications. Furthermore, he describes three groups of advices according to their properties. \emph{Spectative} aspects, which do not influence the underlying computation, \emph{Regulatory} aspects, which change the control flow  but do not affect existing fields, and \emph{Invasive} aspects, which affect existing fields. This classification is similar to ours, however, our characterization of is more fine grained. The two first correspond to our behavioral classification and the last to our data access classification. 

	Clifton and Leavens propose \emph{Spectators} and \emph{Assistants}~\cite{clifton:aosd-foal02}. Spectators are advices that do not affect the control flow of the advised method and do not affect existing fields. Assistants can change the control flow  of the advised method and affect existing fields. \emph{Spectator} are similar to our classification category \emph{Augmentation}  and \emph{Read} in the sense that they do not interfere with the mainline computation or write fields. All other classification categories are equivalent to \emph{Assistants}. Nevertheless, we have achieved a more fine granularity level in our classification.

	\section{Conclusions and Future work}

In this report we have presented a characterization of invasive behavior. Such a characterization is encoded as a classification of invasive aspect oriented programing. The component of this classification represent invasiveness patterns, i.e. the ways in which AOP can be invasive. 

The characterization of invasive aspects allow developers to recognize and reason about the potential risks introduced by aspects into the base program. We think that this is the first step to a bigger specification framework that will make developers more confident in AOP. In future work we will explore the creation of an aspects - base program specification framework.
	
	\bibliographystyle{plain} 
	\bibliography{bibliography}
	\newpage
	\tableofcontents

\end{document}